\documentclass[aps,prapplied, twocolumn,showpacs,amsmath,amssymb,superscriptaddress]{revtex4-2}
\usepackage{hyperref}
\usepackage{graphicx}
\usepackage{dcolumn}
\usepackage{bm}
\usepackage{caption}
\usepackage{upgreek}
\usepackage[utf8]{inputenc}
\usepackage[T1]{fontenc}
\usepackage{mathptmx}
\usepackage{amsmath}
\usepackage{float}
\usepackage{array}
\usepackage{gensymb}
\usepackage[section]{placeins}
\newcommand{\graX}[3][1]{
	\begin{figure}
		\includegraphics[width=#1\columnwidth]{#2}
		\caption{#3}
	\end{figure}
    \vskip 10pt
}

\makeatother
\begin{document}
	\title{Two-color grating magneto-optical trap for narrow-line laser cooling}
	
	\author{S. Bondza}
	\email{saskia.bondza@ptb.de}
	\affiliation{Physikalisch-Technische Bundesanstalt, Bundesallee 100, 38116 Braunschweig, Germany}
	\affiliation{Deutsches Zentrum für Luft- und Raumfahrt e.V. (DLR), Institut f\"ur Satellitengeod\"asie und Inertialsensorik, c/o Leibniz Universit\"at Hannover, Callinstraße 36, 30167 Hannover, Germany}
	\author{C. Lisdat}
	\email{christian.lisdat@ptb.de}
	\affiliation{Physikalisch-Technische Bundesanstalt, Bundesallee 100, 38116 Braunschweig, Germany}
	\author{S. Kroker}
	\affiliation{Technische Universit\"at Braunschweig, Institut für Halbleitertechnik, Hans-Sommer-Str. 66, 38106 Braunschweig, Germany}
	\affiliation{LENA Laboratory for Emerging Nanometrology, Langer Kamp 6, 38106 Braunschweig, Germany}
	\affiliation{Physikalisch-Technische Bundesanstalt, Bundesallee 100, 38116 Braunschweig, Germany}
	\author{T. Leopold}
	\affiliation{Physikalisch-Technische Bundesanstalt, Bundesallee 100, 38116 Braunschweig, Germany}	
	\affiliation{Deutsches Zentrum für Luft- und Raumfahrt e.V. (DLR), Institut f\"ur Satellitengeod\"asie und Inertialsensorik, c/o Leibniz Universit\"at Hannover, Callinstraße 36, 30167 Hannover, Germany}

	\date{\today}

\begin{abstract}
We demonstrate for the first time the two-color cooling and trapping of alkaline-earth atoms in a grating magneto-optical trap (gMOT). The trap is formed by a single incident laser beam together with four secondary beams that are generated via diffraction from a nanostructured wafer. A grating structure for a gMOT operating with strontium atoms is optimized and fabricated. We trap $10^6$ $^{88}$Sr atoms on the $^1$S$_0$ $\rightarrow$ $^1$P$_1$ transition at $461\;\mathrm{nm}$ and transfer $25\;\%$ of these atoms to the second cooling stage on the narrower $^1$S$_0$ $\rightarrow$ $^3$P$_1$ intercombination transition at $689\;\mathrm{nm}$, preparing a sample of $2.5\times 10^5$ atoms at $5\;\mu$K. These results demonstrate for the first time the applicability of the gMOT technology in conjunction with two widely differing wavelengths and enable the continued miniaturization of alkaline-earth based quantum technologies like optical atomic clocks.
\end{abstract}

	\maketitle

\section{Introduction}
Laser-cooled and trapped atoms provide a successful technological basis for second-generation quantum sensors. Among these devices are highly sensitive magnetometers \cite{wil05d,coh19}, gravimeters \cite{kas92,deb11,abe16}, quantum computers \cite{saf16,hen20} and optical atomic clocks \cite{lud15}. All these devices experience a transition from a laboratory to field-based environment \cite{gro18a, ohm21, fan16, lin19} and a beginning commercialization \cite{men18a}, which requires a high degree of miniaturization and integration of the basic enabling technologies.

Compact and scalable devices for cooling and trapping of atoms at a single wavelength have recently been demonstrated in the form of grating magneto-optical traps (gMOTs) \cite{van10b,nsh13,mcg17,bar19b,sit21}, where one incident laser beam produces secondary beams via diffraction on a set of gratings. Combined, these beams lead to cooling and confinement of the atoms above the gratings when used in conjunction with a suitable magnetic quadrupole field. Advanced atom interferometers, as well as optical atomic clocks based on cold atoms benefit from an ultra-narrow electronic transition \cite{hu17a, lud15}, which is provided for example in alkaline-earth-like atoms. However, due to the broad linewidth of the $^1$S$_0$ $\rightarrow$ $^1$P$_1$ transition in alkaline-earth atoms and the correspondingly high Doppler temperature, additional cooling on the narrower $^1$S$_0$ $\rightarrow$ $^3$P$_1$ transition is commonly applied to reach temperatures in the low $\mu$K range. The wavelength ratio of these two transitions is typically about 1.5, which leads to significant deviations of the respective diffraction properties on a grating. The difference in diffraction angle leads to a spatial shift of the trapping volumes complicating transfer from the first to the second cooling stage. Furthermore, in a gMOT design it is generally desirable to have a low 0$^{\mathrm{th}}$-order diffraction efficiency for the incident light, which is often achieved by choosing a grating height that corresponds to a quarter of the incident wavelength \cite{mcg16,cot16,bar19b} and cannot be fulfilled for both wavelengths simultaneously. Additionally, the optimum diffraction efficiency in 1$^{\mathrm{st}}$-order is fixed by the number of grating segments and should be identical for transverse electric (TE) and transverse magnetic (TM) polarization components. It is easy to see that the limited number of parameters in grating design do not allow an optimization of the full set of diffraction properties, especially for widely varying wavelengths.

In this paper we present a grating structure dedicated to magneto-optical trapping of strontium atoms in a gMOT and demonstrate the first two-stage cooling of alkaline-earth atoms in such a system. We enhance the optimization parameter space by applying a thin gold coating on top of the grating, which allows us to balance the diffraction efficiencies for 461 nm and 689 nm. A central hole in the wafer removes the relevance of the 0$^{\mathrm{th}}$-order diffraction efficiency for the second cooling stage at 689 nm, as the atoms have coalesced to a cloud of millimeter size at the end of the first cooling stage. 

With a grating chip fabricated according to our design, we prepare $2.5\times 10^5$ atoms at a temperature of 5 $\mu$K, which is sufficiently low for a transfer into a typical optical lattice as used in atomic clocks \cite{fal14,pol14}. We describe the transfer process from the first to the second cooling stage with an achieved transfer efficiency of $25\;\%$.
\section{Grating Design}

\begin{figure*}
	\captionsetup{type=figure}
	\includegraphics[width=0.95 \textwidth]{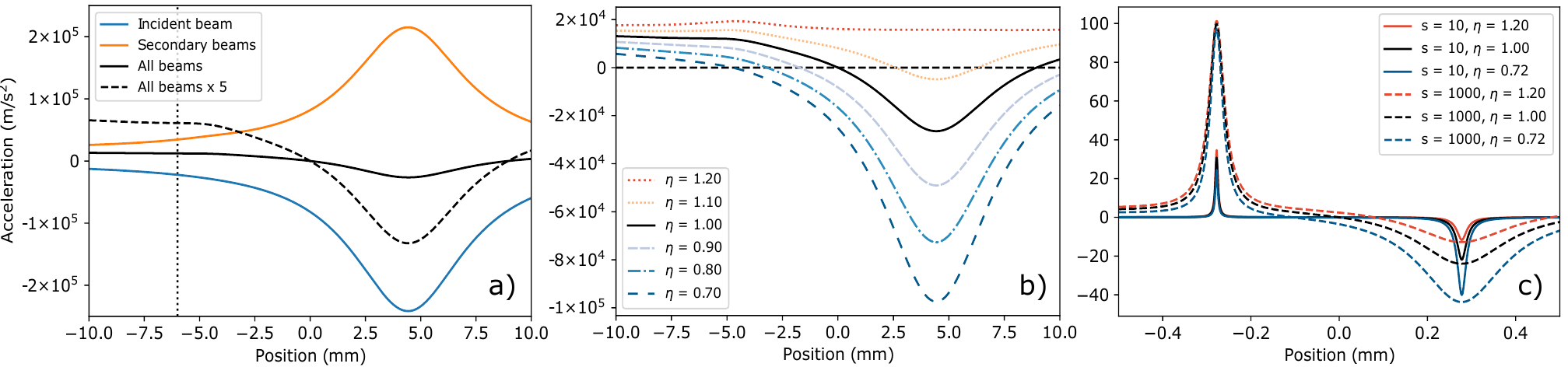}
	\captionof{figure}{Illustration of axial acceleration acting on an atom at rest along the symmetry axis $z$ of the quadrupole field for both cooling stages. The magnetic field minimum marks zero on the $z$-axis. a) Contribution of incident and secondary beams to the axial acceleration in the first stage MOT with a saturation parameter $s=0.5$, a detuning of $-32\;\mathrm{MHz}$, magnetic field gradient $\frac{dB}{dz}=5\;\mathrm{mT/cm}$ and diffraction angle $\alpha_{461} = 31\degree$. The dotted vertical line marks the position of the grating. b) Influence of intensity imbalance, characterized by $\eta$, on the total axial acceleration in the first stage MOT. Imbalance $(\eta\neq 1)$ leads to a shift of the steady state position, as seen by the intersection of the acceleration curve with the zero acceleration line (dashed black line). c) Total axial acceleration for different saturation parameters and levels of intensity balance in the second stage MOT. The case $\eta=0.72$ corresponds to the manufactured grating, see Table \ref{tab: diffraction eff}. $ \Delta = -400~\mathrm{kHz}$, $\frac{dB}{dz} = 0.9~\mathrm{mT/cm}$ and $\alpha_{689} = 50\degree$. \label{fig: Forces in MOT}}
\end{figure*}
The idea of restoring forces in a magneto-optical trap (MOT) is most easily realized by laser beams incident along the $x$, $y$-, and $z$-axis of the quadrupole field $\mathbf{B} = b\{x,y,-2z\}$ with circular polarization. Choosing the correct handedness the $\sigma$ transitions are predominantly driven which leads to a restoring force. In a gMOT this is not the case, as the secondary beams form an oblique angle with all principal axes. Hence, coupling of the secondary beams to all ($\sigma^+$,$\pi$,$\sigma^-$) transitions must be evaluated \cite{van11b, van09}. Looking at the laser-induced acceleration of an atom at rest along the symmetry axis $z$ of the quadrupole field, we see that the acceleration from the secondary beams exhibits a trapping and an anti-trapping contribution due to coupling to all $\Delta m=0,\pm 1$ transitions as illustrated in Fig. \ref{fig: Forces in MOT} \cite{van11b}. Nevertheless, the figure shows that stable trapping is possible as the combined acceleration of incident and secondary beams can lead to an overall restoring force, depending on the choice of grating parameters.

Here, we present our design of a grating structure for two-stage cooling of strontium atoms. The first stage utilizes the $^1$S$_0$ $\rightarrow$ $^1$P$_1$ transition with a linewidth of $30.2\;\mathrm{MHz}$ at a wavelength of $461\;\mathrm{nm}$ \cite{nic15} and the second stage the $^1$S$_0$ $\rightarrow$ $^3$P$_1$ transition with a linewidth of $7.5\;\mathrm{kHz}$ at a wavelength of $689\;\mathrm{nm}$ \cite{yas06}. The second stage is subdivided into a broadband (BB) phase, where the laser linewidth is artificially broadened to the megahertz range, and a single-frequency (SF) phase, where the laser operates with a linewidth on the order of the transition linewidth. This procedure increases the transfer efficiency of atoms from the first stage to the second cooling stage \cite{kat99}. When designing a grating for two very different wavelengths one should mind several considerations that are connected to the wavelength dispersion of the diffraction angle and diffraction efficiency: 

(I) Differences in diffraction angle result in an unfavorable axial shift and therefore only partial overlap of trapping volumes. The trapping volume for each wavelength is defined as the volume in space where light from both the primary and all secondary beams of the respective color is present, see Fig. \ref{fig:twocolourMOT}. A small overlap of both volumes complicates transfer of the atoms from the first to the second stage MOT. 

(II) A diffraction efficiency of $1/N$ for both colors, where $N$ is the number of secondary beams, leads to intensity balance \cite{van10b}, characterized by a vanishing net force in the absence of a magnetic field i.e. the intensity-weighted k-vectors of all beams adding to zero. Additionally, for both trapping wavelengths equal diffraction efficiencies of the TE and TM polarization mode are important in order to conserve the incident circular polarization, as radial trapping weakens otherwise.

(III) The 0$^{\mathrm{th}}$-order diffraction for both wavelengths should be minimized, as it contributes predominantly to an anti-trapping acceleration of the atoms \cite{cot16}.

(IV) Varying the grating period shows three effects. First of all, the overlap of the trapping volumes increases for larger grating periods. Secondly, larger diffraction angles are obtained for shorter grating periods resulting in increased radial and axial restoring forces. At last, a geometry for simultaneous optimization of 1$^{\mathrm{st}}$-order diffraction efficiencies to about $1/N$ for both wavelengths and polarization modes cannot be found for all grating periods.

(V) Intensity balance is of particular importance for magneto-optical trapping on broad transitions, when the maximum achievable magnetic field gradient is not sufficient to enable spatially well-resolved addressing of the transitions. The effect of deviations from optimum diffraction efficiency are shown in Fig. \ref{fig: Forces in MOT}b). Here, deviations from intensity balance, characterized by the figure of merit $\eta$, which is the ratio of the diffraction efficiency to its ideal value of $1/N$, result in a shift of the equilibrium position in the trap with respect to the magnetic field minimum and a reduced trap depth. In contrast, axial cooling on the narrow transition can actually benefit from intensity imbalance when cooling with laser intensities $I$ well above saturation ($s=I/I_{\mathrm{sat}}>>1$). As shown in Fig. \ref{fig: Forces in MOT}c), the trap depth increases for $\eta<1$ due to a decrease of the anti-trapping contribution of the secondary beams relative to the trapping contribution of the incident beam. The trapping contribution of the secondary beams is not significantly reduced due to saturation.

With these constraints in mind, we designed a grating structure which is optimized for two-stage magneto-optical trapping of strontium atoms. The 0$^{\mathrm{th}}$-order diffraction is geometrically minimized for $461\;\mathrm{nm}$, while for $689\;\mathrm{nm}$ a $3\;\mathrm{mm}$ hole in the center of the chip suppresses back reflection in the relevant volume for second-stage trapping. The parameters for optimization are grating period, height, fill factor and the thickness of an additional gold layer. Varying the fill factor for a fixed grating period and heigth allows to change the ratio of TE to TM diffraction efficiency \cite{cot16,mcg17}. The gold layer was introduced to reduce the diffraction efficiency for $461\;\mathrm{nm}$ relative to $689\;\mathrm{nm}$ by means of the higher absorption of gold in the blue spectrum.

One can then devise a cost function that contains the relative deviations of all diffraction efficiencies from their target value and weigh the individual contributions by priority. Optimized grating parameters are found by evaluating the cost function with a simulation program for different parameter sets. We have used rigorous coupled wave analysis (RCWA) \cite{moh83} to scan grating height and duty cycle for various grating periods and thicknesses of the gold layer for both silver and aluminium substrates. A four-segment design sets the targeted 1$^{\mathrm{st}}$-order diffraction efficiency to $25\;\%$ and enabled lower cost functions compared to a three-segment design. Best results were obtained for an $900\;\mathrm{nm}$ period aluminium grating corresponding to diffraction angles of $\alpha_{461}=31\degree$ and $\alpha_{689} = 50\degree$ for $461\;\mathrm{nm}$ and $689\;\mathrm{nm}$ respectively. The optimum height of the grating grooves was evaluated to $109\;\mathrm{nm}$ with a fill factor of 0.283. The structure is coated with a $10\;\mathrm{nm}$ gold layer. 
The grating was manufactured in the PTB clean room facilities. A silicon wafer was coated with a base layer of $100\;\mathrm{nm}$ aluminium. The grating lines were produced using electron beam lithography and a lift-off process. The gold layer was evaporated on top of the structure.

The diffraction efficiencies were measured experimentally and are in general agreement with simulated values as presented in Table \ref{tab: diffraction eff}. The ratio of the 1$^{\mathrm{st}}$-order diffraction efficiency for TE and TM polarization mode is $\geq 0.89\;\%$ at both wavelengths. The grating geometry was inspected with atomic force microscopy, scatterometry and scanning electron microscopy. Good agreement between design and fabricated geometry suggests that deviations between measurement and simulation are most likely explained by uncertainties in the material dispersion properties of the thin-film layers and surface oxidation between processing steps.

\begin{table}
	\centering
	\begin{tabular}{c c c c c} 
		\hline \hline
		&\multicolumn{2}{c}{0$^{\mathrm{th}}$ order} 
		& \multicolumn{2}{c}{1$^{\mathrm{st}}$ order} \\
		& simulated  & measured &  simulated&  measured\\ 
		\hline 
		$461\;\mathrm{nm}$ & $8.4$ & $7(1)$ &$26.6$& $25(1)$\\ 
		
		$689\;\mathrm{nm}$ & $46.6$ & $50(2)$ & $21.4$& $18(1)$ \\
		
		\hline \hline
	\end{tabular} 
	\caption{Diffraction efficiencies (averaged over TE and TM modes) of the grating in percent as designed and as measured for $461\;\mathrm{nm}$ and $689\;\mathrm{nm}$. The measured intensity imbalance factor $\eta$ equals unity and 0.72 for first and second cooling stage, respectively. \label{tab: diffraction eff}}
\end{table}

In the second stage MOT, the maximum optical acceleration is only 10 times stronger than gravity such that the effect of the latter can no longer be neglected, see Fig. \ref{fig: Forces in MOT}c). The SF MOT forms at a height where gravity is balanced by optical acceleration so that trapped atoms will gather on the flank of the acceleration peak opposing gravity. In a micro-gravity environment the atoms will scatter on both acceleration peaks.

In the direction opposing the incident beam, the acceleration results from the difference of a large trapping and a slightly smaller anti-trapping light force, analog to the first cooling stage, see Fig. \ref{fig: Forces in MOT}a). Choosing the equilibrium position on that scattering peak thus leads to an overall increase in scattering events and, hence, a higher temperature in the MOT is expected compared to the gravitationally inverted situation or a six-beam MOT. In this work, we operate in the above described "worst case" geometry, with gravity pointing away from the chip surface to prove the general feasibility of second stage cooling of strontium in a gMOT, independent of orientation. 

\section{Results}
The grating chip is mounted inside an ultra-high vacuum system comprised of a standard spherical octagon and two vacuum pumps, an $80\;\mathrm{l/s}$ non-evaporable getter pump and a $40\;\mathrm{l/s}$ ion pump on opposite sides of the vacuum system. For the generation of a quadrupole field with variable magnetic field gradients up to $6\;\mathrm{mT/cm}$, two wire coils are mounted within the vacuum system. Their symmetry plane lies $6\;\mathrm{mm}$ above the grating surface. Additional coils outside the vacuum system allow fine-positioning of the magnetic field minimum. 

\graX[0.95]{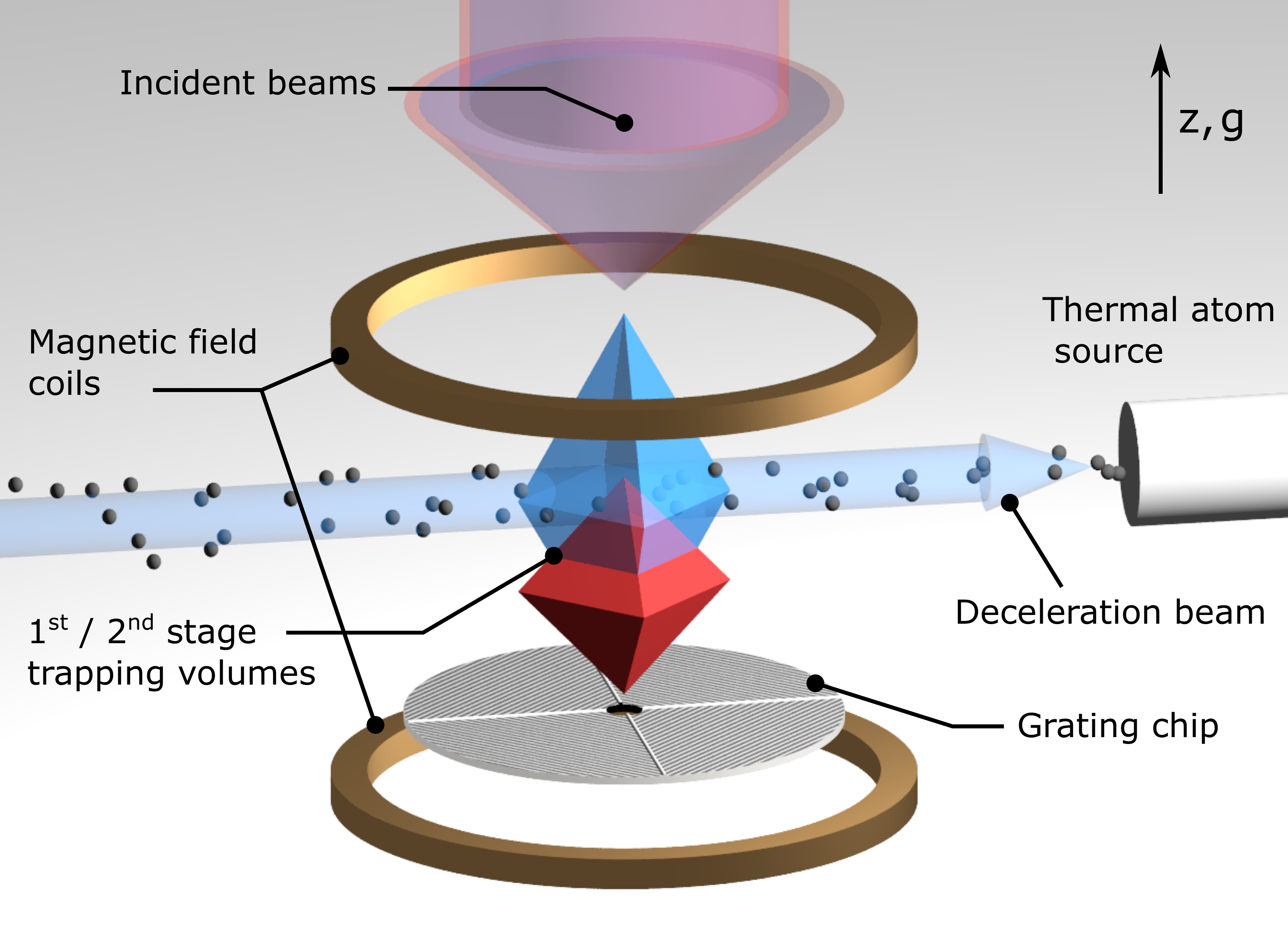}{Schematical representation of the experimental setup. $461\;\mathrm{nm}$ and $689\;\mathrm{nm}$ laser beams with $18\;\mathrm{mm}$ diameter are incident on the grating chip, producing secondary beams by diffraction. The volumes where light from the incident and each secondary beam overlap form an octahedron above the chip, where the elongation is determined by the diffraction angle. Magnetic field coils are mounted symmetrically to a point in the overlap of the trapping volumes. Radially incident atoms from a thermal source are slowed by a counterpropagating deceleration beam. \label{fig:twocolourMOT}}

Atoms are evaporated from a thermal source following \cite{sch12c}. In contrast to most other experiments trapping strontium atoms, we do not employ a Zeeman slower between the thermal source and our MOT. Atoms are decelerated by a slowing beam, red-detuned by $220\;\mathrm{MHz}$ with respect to the $^1$S$_0$ $\rightarrow$ $^1$P$_1$ transition as depicted in Fig. \ref{fig:twocolourMOT}. The lower fraction of slow atoms compared to a Zeeman-slowed beam is partially compensated by a short distance between the atom source and the trapping region of only $14\;\mathrm{cm}$. With the deceleration beam turned off, the MOT loading rate is reduced by a factor of $>100$.

The base vacuum pressure in our system is $2 \times 10^{-10}\;\mathrm{mbar}$ and rises up to $5\times 10^{-9}\;\mathrm{mbar}$ for temperatures of the strontium oven up to $390\;\degree\mathrm{C}$.

A beam shaping unit consisting of a variable magnification telescope and a flat-top beam shaper combines fiber-coupled light at $461\;\mathrm{nm}$ and $689\;\mathrm{nm}$, producing an $18\;\mathrm{mm}$ diameter bi-chromatic laser beam with circular polarization and a flat-top intensity profile. The grating structure is illuminated with maximum laser intensities corresponding to $0.5\;I_{\mathrm{sat}}$ at $461\;\mathrm{nm}$ and $2000\;I_{\mathrm{sat}}$ at $689\;\mathrm{nm}$, respectively.

\graX[0.9]{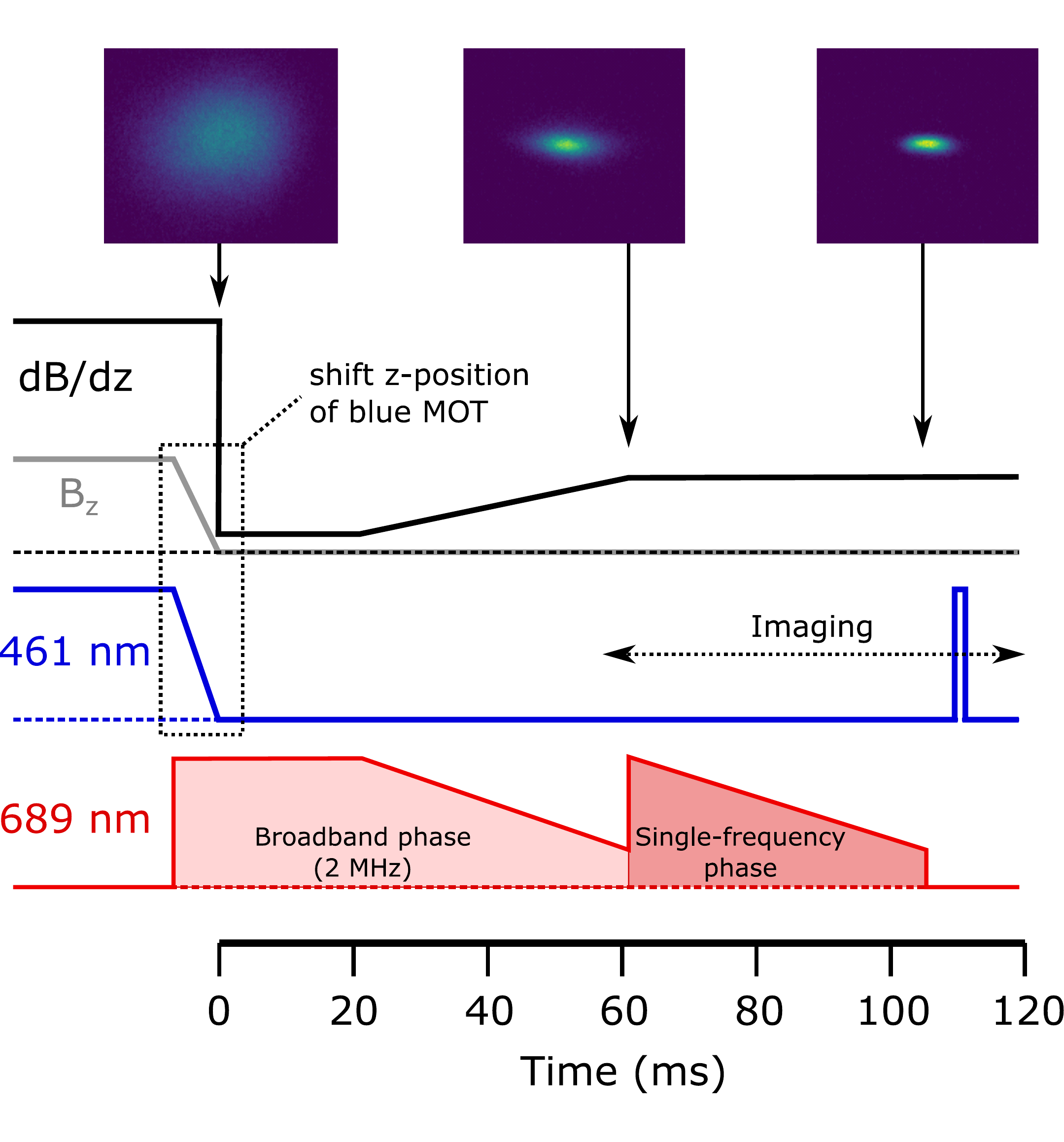}{Time evolution of experimental parameters: Axial magnetic field gradient (black), axial magnetic offset field (grey), $461\;\mathrm{nm}$ laser intensity (blue) and $689\;\mathrm{nm}$ laser intensity (red). Fluorescence images of the MOT at different times during the cooling sequence are shown. For details see information in the text. \label{fig:Ex_Seq}}

The $^1$S$_0$ $\rightarrow$ $^1$P$_1$ transition is not closed due to decay from the $^1$P$_1$ state to the $^1$D$_2$ state and subsequent decay to the $^3$P$_2$ state. To close this loss channel we employ a standard repump scheme outlined in \cite{fal11} using two additional lasers with wavelengths of $679\;\mathrm{nm}$ and $707\;\mathrm{nm}$. Both wavelengths are combined using a dichroic mirror and illuminate the atoms radially. Another radial axis is used for a laser beam resonant with the $^1$S$_0 \rightarrow ^1$P$_1$ transition at $461\;\mathrm{nm}$ for fluorescence detection. 

The light sources for $461$, $679$ and $707\;\mathrm{nm}$ are commercial external cavity diode lasers which are stabilized to a wavemeter. Narrow-linewidth laser light at $689\;\mathrm{nm}$ is produced with a home-built laser system based on self-injection locking \cite{lin12a} of a single-frequency laser diode (Eblana Photonics). Long-term frequency stability is achieved by referencing the laser to a vacuum-housed optical resonator. We estimate a short-term laser linewidth of $50\;\mathrm{kHz}$ from the in-loop error signal of the feedback loop to the resonator. A high power Fabry-P\'erot laser diode is injection-locked from this laser system yielding up to $15\;\mathrm{mW}$ of light incident on the grating.

The fluorescence light scattered by the atoms from the detection beam is imaged parallelly to the grating surface onto either a camera or photomultiplier tube with a magnification of 0.8. 

The experimental sequence to cool atoms to $5\;\mathrm{\mu \mathrm{K}}$ in the SF MOT is depicted in Fig. \ref{fig:Ex_Seq}. It starts with loading about $10^6$ atoms into the first cooling stage. This is done by shining in the $18\;\mathrm{mm}$ diameter red-detuned first-stage cooling beam onto the grating and the deceleration beam onto the atom source. The diameter of the cooling beam results in a maximum radial extension of the first-stage trapping volume at a height of $7.5\;\mathrm{mm}$ above the chip surface, $1.5\;\mathrm{mm}$ above the symmetry point of the \textit{in vacuo} magnetic field coils. A quadrupole magnetic field with a gradient of $5\;\mathrm{mT/cm}$ along the $z$ axis is produced with the wire coils. An external offset magnetic field is applied along the $z$ direction, to move the magnetic field zero of the quadrupole field towards the symmetry point of the trapping volume. After loading atoms for $300\;\mathrm{ms}$ the deceleration beam is turned off, removing the additional radiation pressure and changing the MOT position slightly. The second-stage cooling laser is turned on at full power, with a laser linewidth broadened to $1.6\;\mathrm{MHz}$ by frequency modulation at $50\;\mathrm{kHz}$. The external magnetic offset field and the first-stage cooling laser power are ramped to zero. This moves the atoms towards the grating chip, inside the second-stage trapping volume, see Fig. \ref{fig:twocolourMOT}. After moving the atoms, the magnetic field gradient is switched to $0.3\;\mathrm{mT/cm}$ to reduce the Zeeman shift to within the laser bandwidth. After cooling for $15\;\mathrm{ms}$ the BB MOT is spatially compressed by linearly increasing the magnetic field gradient up to $1\;\mathrm{mT/cm}$. At the same time, the laser intensity is reduced to reach lower temperatures.

The SF MOT phase starts by turning off the artificial laser linewidth broadening and increasing the laser intensity to the maximum value of $s=2000$. Over the course of several $10\;\mathrm{ms}$, the intensity is reduced. In the fluorescence images in Fig. \ref{fig:Ex_Seq}, a further compression of the atoms in the SF phase compared to the BB phase is visible. We were able to trap $10^6$ atoms in the first cooling stage and achieve a transfer efficiency of $25\;\%$ to the SF MOT, corresponding to $2.5\times 10^5$ atoms.

We measure the lifetime in the first cooling stage as a function of magnetic field gradient and laser intensity. Lifetime measurements were performed by loading the MOT, turning off the deceleration beam, which effectively stops loading atoms, and measuring the relative number of atoms after a variable wait time before detection. 

We found the lifetime to increase with magnetic field gradient and laser intensity as depicted in Fig. \ref{lifetime}, similar as in \cite{sit21}. We measured a lifetime of up to $700\;\mathrm{ms}$ with our maximum available laser intensity and magnetic field gradient at a pressure of $3\times 10^{-9}$ mbar which is close to being vacuum limited \cite{nag03}.

\graX{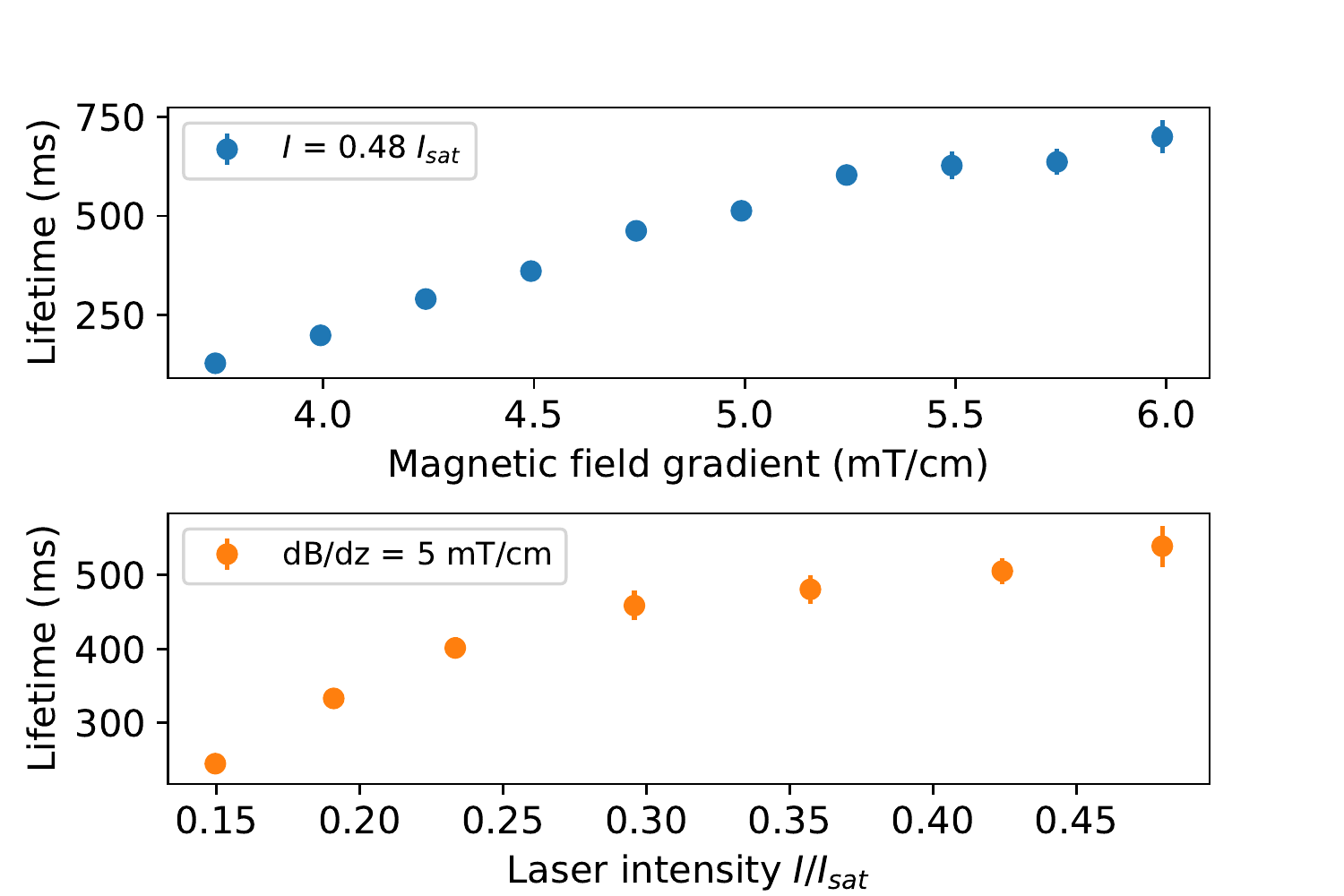}{Lifetime in the first stage MOT in dependence of laser intensity and magnetic field gradient. Error bars are about the size of the data points. \label{lifetime}} 
The necessity for a relatively high laser intensity can be traced back to the reduced trap depth in the gMOT configuration. From Fig. \ref{fig: Forces in MOT}a) one can infer that the total axial acceleration, and hence trap depth, for a given laser intensity is about 10 times smaller than in a conventional six-beam MOT, where the magnitude of restoring acceleration is roughly the size of the contribution from the incident beam. An increase in trap depth with magnetic field gradient is expected as long as the Zeeman shift remains smaller than the laser detuning within the trapping volume \cite{yoo07a}.

We characterize the MOT temperature via the time-of-flight expansion method using fluorescence imaging with $100\;\mu\mathrm{s}$ stroboscopic illumination from the detection beam. The size of the atom cloud $\sigma$ is then plotted as a function of time $t$ and fitted according to $\sigma^2(t)=\sigma_0 ^2 + \frac{k_B T}{M}\cdot t^2$, where $\sigma_0$ is the initial size of the atom cloud, $k_B$ denotes Boltzmann's constant, $M$ the atomic mass and $T$ the atomic temperature. We fit separate temperatures for the axial and radial projection of the MOT onto the camera.

\graX[1]{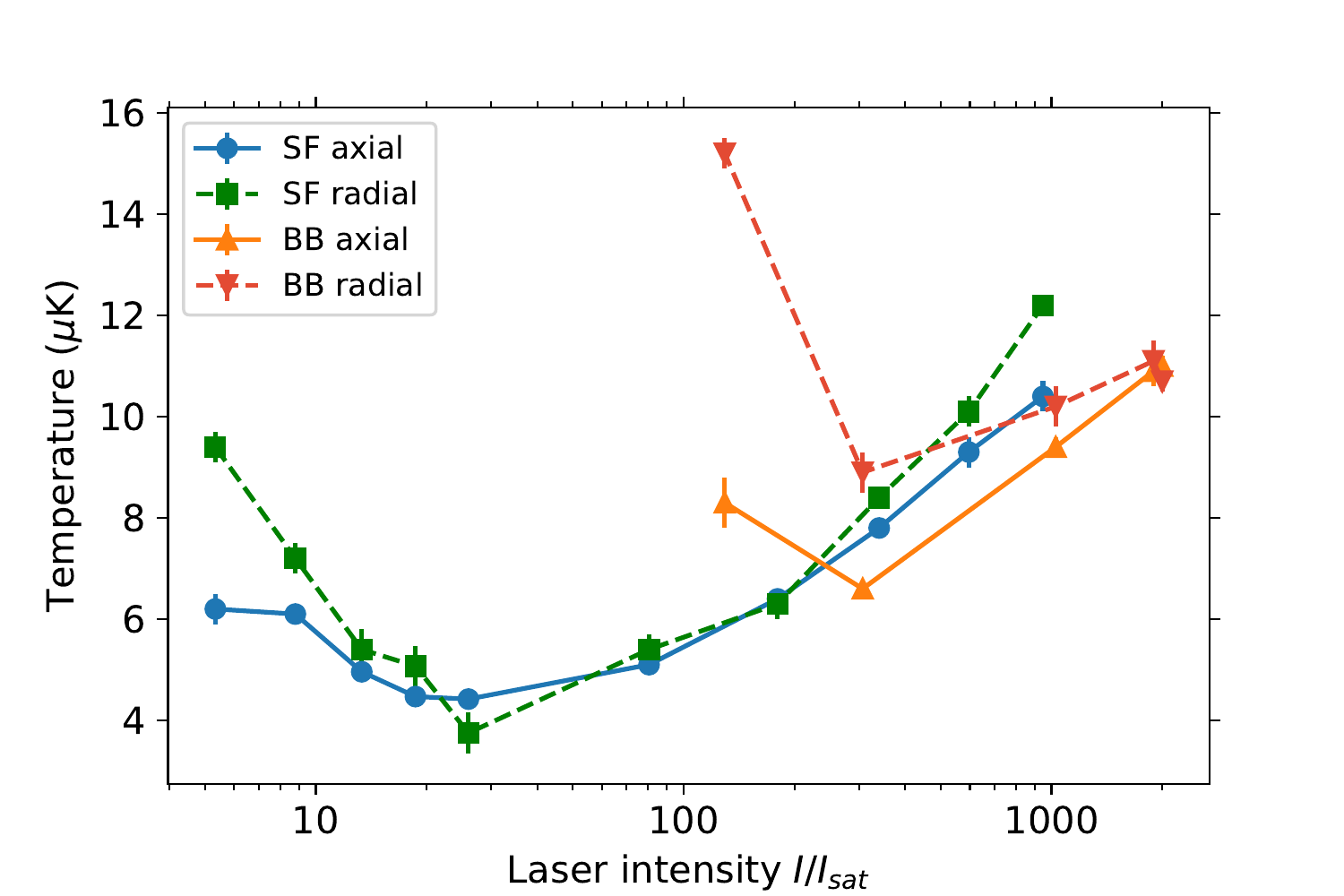}{Temperature in the BB and SF MOT in radial and axial direction as function of laser intensity. \label{fig:red_Mot_temperature}}
We find the temperature in the first stage MOT to increase with increasing intensity of the laser beams as is generally expected \cite{xu03}. At $s=0.5$ we measure an axial temperature of $3.5-4\;\mathrm{mK}$ and a radial temperature of $7.5-8\;\mathrm{mK}$. Lowering the intensity at the end of the cooling cycle, we reach temperatures of $1.2\;\mathrm{mK}$ and $4.5\;\mathrm{mK}$ in axial and radial direction, respectively. In a conventional six-beam configuration temperatures of $1\;\mathrm{mK}$ can be achieved with these parameters \cite{xu03}. Higher temperatures can qualitatively be explained by the anti-trapping contribution from the secondary beams.

We measure the temperature in the second stage BB and SF MOT as a function of the incident laser intensity. In the BB MOT, temperatures of around $10\;\mu\mathrm{K}$ are reached as seen in Fig. \ref{fig:red_Mot_temperature}, which is comparable to the standard six-beam configuration \cite{pol14}. In the SF MOT we reach minimum temperatures of slightly below $5\;\mu\mathrm{K}$ both in radial and axial direction. We find the temperature to decrease with laser intensity, as expected, until it reaches a minimum at around $20-30$ I$_{\mathrm{sat}}$. For lower laser intensities the MOT temperature increases again. In a six-beam geometry temperature in the SF MOT decreases continuously with intensity down to sub-$\mu$K \cite{lof04}. We suspect that the temperature is limited by the anti-trapping interaction with the secondary beams, as discussed earlier, and expect that lower temperatures can be reached by mounting the grating such that the secondary beams counteract gravity, which suppresses interaction with the primary beam and thus reduces the scattering rate in the SF MOT.

\section{Discussion}
We have demonstrated magneto-optical trapping of alkaline-earth atoms in a gMOT on both the $^1$S$_0$ $\rightarrow$ $^1$P$_1$ and the narrow-line $^1$S$_0$ $\rightarrow$ $^3$P$_1$ cooling transition. This is the first time that sequential magneto-optical trapping on two transitions with different wavelengths is shown in a gMOT. We prepare $2.5\times10^5$ $^{88}$Sr atoms at a temperature of $5\;\mu\mathrm{K}$ which is sufficiently cold for loading into an optical lattice used in optical clocks, with typical trap depths on the order of $10\;\mu\mathrm{K}$. We have given general guidelines for the design of a grating for two-colour magneto-optical trapping and expect that the results of our work can be transferred to other atomic species that require sequential cooling stages, such as ytterbium.

We expect that the number of trapped atoms can be increased significantly by a higher flux of slow atoms using a Zeeman slower. Differential pumping between the atom source and science chamber would improve the vacuum level and reduce the loss rate from the MOT. Additionally, higher loading rates can be achieved in a gMOT by loading atoms along the axial direction of the grating \cite{bar19b}. To improve our understanding of the single-beam geometry for cooling on narrow linewidth transitions we plan an in-depth study based on Monte-Carlo simulations.

Our results prove that the gMOT technology is compatible with multi-color magneto-optical trapping. This enables highly compact quantum sensors based on alkaline-earth-like atoms, e.g. optical atomic clocks and atom interferometers.
\begin{acknowledgments}
We thank Kathrin St\"orr and Thomas Weimann for the fabrication of the grating chip and Carsten Feist for laser cutting of the wafer. We would further like to thank Frank Fuchs / Gitterwerk GmbH for providing the RCWA code Moose. 
This work was financially supported by the State of Lower-Saxony through the VW Vorab and DLR, project D/123/67284017. We further acknowledge support by the 
Deutsche Forschungsgemeinschaft (DFG, German Research Foundation) under Germany’s Excellence Strategy -- EXC-2123 QuantumFrontiers -- Project-ID 390837967 and SFB~1464 TerraQ -- Project-ID 434617780 -- within project A04. 
\end{acknowledgments}

\bibliography{gMOTdraft_twocolor_Strontium}

\end{document}